%%%%%%%%%%%%%%%%%%%%%%%%%%%%%%%%%%%%%%%%%%%%%%%%%%%%%%%%%%%%%%%%%%%%%%
%                                                                    %
% Noncommutative Field Theory on Yang's Space-time Algebra, Covariant%
% Moyal Star Product and Matrix Model                                %                                                                    %
%                                                                    %
%                     by Sho Tanaka                                  %
%                     TeX with PHYZZX macro                          %
%%%%%%%%%%%%%%%%%%%%%%%%%%%%%%%%%%%%%%%%%%%%%%%%%%%%%%%%%%%%%%%%%%%%%%
\input phyzzx.tex
\hoffset=10mm
\vsize=26cm
\titlepage 
\title{{\bf Noncommutative Field Theory on Yang's Space-Time Algebra, Covariant Moyal 
Star Product and Matrix Model} }
\author{Sho TANAKA\footnote{*}{Em.~Professor of Kyoto University, E-mail: stanaka@yukawa.kyoto-u.ac.jp}}
\address{Kurodani 33-4, Sakyo-ku, Kyoto 606-8331, Japan}
\abstract{Noncommutative field theory on Yang's quantized space-time algebra (YSTA) 
is studied. It gives a theoretical framework to reformulate the matrix model as 
quantum mechanics of $D_0$  branes in a Lorentz-covariant form. The so-called kinetic term 
($\sim {\hat{P_i}}^2)$ and potential term ($\sim {[\hat{X_i},\hat{X_j}]}^2)$ of $D_0$ branes 
in the matrix model are described now in terms of Casimir operator of $SO(D,1)$, a 
subalgebra of the primary algebra $SO(D+1,1)$ which underlies YSTA with two contraction- 
parameters, $\lambda$ and $R$. $D$-dimensional 
noncommutative space-time and momentum operators $\hat{X_\mu}$ and $\hat{P_\mu}$ in                      
YSTA show a distinctive spectral structure, that is, space-components $\hat{X_i}$ and 
$\hat{P_i}$ have discrete eigenvalues, and time-components $\hat{X_0}$ and $\hat{P_0}$ 
continuous eigenvalues, consistently with Lorentz-covariance. According to the method of 
Lorentz-covariant Moyal star product proper to YSTA, the field equation of $D_0$ brane 
on YSTA is derived in a nontrivial form beyond simple Klein-Gordon equation, which reflects 
the noncommutative space-time structure of YSTA.} 

\endpage

\chapter{Introduction}

Along with the development of the matrix model of M-theory,\Ref\M\bfss a lot of field          
theories on noncommutative space-time variables have been proposed.\Ref\Mori\Mo\  
In the most of them, the noncommutativity of space-time variables $\hat{x_i}$ are 
simply assumed in the form  
$$
                   [\hat{x_i},\hat{x_j}] = i \theta_{ij},
\eqn\eqA
$$
with an anti-symmetric constant tensor $\theta_{ij}$. Field theories on noncommutative 
space-time are there related to a local field theory on commutative 
space-time by making use of the method of Moyal star product. Under this procedure, 
the propagator of field remains in most cases to be that of the commutative case and the 
noncommutativity of space-time is solely reflected on the appearance of nonlocal 
form-factor coming from Moyal star operation,  $ \exp\ (i / 2\  \overleftarrow{\partder{}
{x_i}}\theta_{ij}\overrightarrow{\partder{}{x_j}})$, in interaction terms among local 
fields. 

At this point, we wonder why the propagator by itself does not reflect the 
noncommutative structure of space-time expressed by \eqA. One easily finds out that 
the situation simply comes from an assumption that the kinetic term of noncommutative 
field $\Phi$ is given by $ \sim \partder{\Phi}{x_\mu} * \partder{ \Phi}{x_\mu} $ under 
Moyal star product mentioned above. This problem will be our important concern in what 
follows. 
\def\bfss{T.~Banks, W.~Fischler, S.~H.~Shenker and L.~Susskind, ``M Theory As A Matrix Model: 
A Conjecture,"  Phys.~Rev.~{\bf D55} (1997) 5112, hep-th/9610043.}
\def\Mo{A.~Connes, M.R.~Douglas and A.~Schwarz, ``Noncommutative Geometry and Matrix Theory: 
Compactification on Tori," JHEP {\bf 02}, 003, hep-th/9711162; S.~Minwalla, M.V.~Raamsdonk 
and N.~Seiberg, ``Noncommutative Perturbative Dynamics," JHEP {\bf 02},020, hep-th/9912072; 
M.~Chaichian, A.~Demichev and P.~Pre{\u s}najder, ``Quantum Field Theory on Noncommutative 
Space-Times and the Persistence of Ultraviolet Divergences,"  Nucl.~Phys. {\bf 567} (2000) 360, 
hep-th/9812180;  R.J.~Szabo, ``Quantum Field Theory on Noncommutative Spaces," Phys.Rev. 
{\bf 378} (2003) 207, hep-th/0109162; M.R.~Douglas and N.A.~Nekrasov, ``Noncommutative Field 
Theory," Rev.Mod.Phys. {\bf 73} (2001) 977, hep-th/0106048; C.E.~Carlson and C.D.Carone, 
``Noncommutative Gauge Theory without Lorentz Violation," Phys.~Rev. {\bf D66} (2002)075001, 
hep-th/0206035; H.~Kase, K.~Morita, Y.~Okumura and E.~Umezawa, ``Lorentz-invariant Non-Commutative 
Space-Time Based On DFR Algebra," hep-th/0212176.}

In the present paper, we wish to extend our previous attempts,\ref\TN \Ref\Ti\TR 
\Ref\Tii\TH\ to construct a noncommutative field theory of the matrix model, which is developed
 on a covariant quantized space-time algebra early proposed by C.N.~Yang\Ref\Ya\yang\ 
immediately after H.S.~Snyder,\Ref\Sn\snyder\ in place of $\eqA$. As was emphasized in 
ref.\Ti, Yang's quantized space-time algebra (YSTA), not simply as a toy model of 
noncommutative space-time algebra, but has several important characteristics. 

First, as was pointed out in ref.\Tii, it should be noticed that, YSTA has a common 
symmetry with $D$-dimensional (Euclidean) CFT, i.e., $SO(D+1,1)$ with two extra 
dimensions. It gives a possibility of regularizing a short distance behavior or 
divergences familiar in  the so-called IR/UV connection in AdS/CFT or dS/CFT 
correspondence, by virtue of discrete spectral structure of space-time and momentum operators 
of YSTA, which CFT lacks entirely.  

\def\TN{S.~Tanaka, ``Space-time quantization and matrix model," Nuovo Cim.{\bf 114 B} (1999) 49, 
hep-th/9808064.}
\def\TR{S.~Tanaka, ``Space-Time Quantization and Nonlocal Field Theory - Relativistic Second 
Quantization of Matrix Model-," hep-th/0002001.}
\def\TH{S.~Tanaka, ``Yang's Quantized Space-time Algebra and Holographic Hypothesis," 
hep-th/0303105.}
\def\yang{C.N.~Yang, Phys.Rev.{\bf 72} (1947) 874; Proc.~of International 
Conf.~on Elementary  Particles, 1965 Kyoto, pp 322-323.}
\def\snyder{H.S.~Snyder, Phys.Rev.{\bf 71} (1947) 38; {\bf 72} (1947),
68.}
\def\VK{N.Ja.~Vilenkin and A.U.~Klimyk, ``Representation of Lie groups and special functions 
II," Kluwer Academic Publishers, 1993.}

Second, as will be shown in the present paper, one finds that YSTA is well fitted for 
covariant description of the matrix model. In fact, the so-called kinetic term 
($\sim {\hat{P_i}}^2$) and potential term ($\sim {[\hat{X_i},\hat{X_j}]}^2)$ of $D_0$ 
branes in the matrix model, which are usually understood to originate in Yang-Mills gauge symmetry, 
are described now in terms of Casimir operator of $SO(D,1)$, a subalgebra of the 
primary algebra $SO(D+1,1)$ which underlies YSTA with two contraction-parameters, $\lambda$ and $R$. 
According to the method of covariant Moyal star product proper to YSTA, we successfully derive 
$D_0$ brane field equation on YSTA in a nontrivial form beyond simple Klein-Gordon equation: 
It clearly reflects the quantized noncommutative space-time structure of YSTA, in contrast 
to most of familiar noncommutative field theories as mentioned in the beginning.

The present paper is organized as follows. In section 2, we shortly recapitulate the algebraic 
structure of Yang's quantized space-time algebra (YSTA) contracted from $SO(D+1,1)$ and 
examine characteristics of the so-called quasi-regular representation,\Ref\QR\VK\ a kind of 
unitary infinite dimensional representation of SO(D+1,1). We investigate translation 
operation in YSTA in connection with momentum operators $\hat{P}_\mu$ and clarify 
how commutative space-time, Heisenberg's uncertainty relation, 
translations and so forth in the ordinary quantum mechanics are restored in YSTA. 
It will be found that they are all together restored in appropriate limiting values of contraction-
parameters 
$\lambda$ and $R$ in a large limit of discrete eigenvalues of reciprocity operator of 
YSTA, which will be called in what follows, the ``quantum-mechanical limit of 
YSTA".

In section 3, we try to rewrite the matrix model covariantly in terms of noncommutative field 
theory on YSTA, by making use of the result in section 2. In order to see a short distance 
behavior of the present noncommutative field theory, we derive the field equation of $D_0$ brane 
by means of covariant Moyal star product proper to YSTA. Preliminary considerations on the new 
equation are given. 

The final section is devoted to discussions and concluding remarks.

\chapter{YSTA and its Quasi-Regular Representation}

 Yang's quantized space-time algebra (YSTA)\refmark\Ya\ was proposed to modify the original 
Snyder's quantized space-time algebra\refmark\Sn\ to be translation-invariant, in addition to 
Lorentz-invariance which holds in both theories. 

$D$-dimensional YSTA is contracted from $SO(D+1,1)$ algebra with generators $\hat{\Sigma}_{MN}$; 
$$
 \hat{\Sigma}_{MN}  \equiv i\ (q_M \partder{}{q_N}-q_N\partder{}{q_M})
\eqn\eqB
$$
which work on $(D+2)$-dimensional parameter space  $q_M$ ($M= \mu,a,b)$ satisfying
$$
             - q_0^2 + q_1^2 + ... + q_{D-1}^2 + q_a^2 + q_b^2 = R^2.
\eqn\eqC
$$ 
Here, $q_0 =-i q_D$ and $M = a, b$ denote two extra dimensions with space-like metric signature.

$D$-dimensional space-time and momentum operators, $\hat{X}_\mu$ and $\hat{P}_\mu$, 
with $\mu =1,2,...,D,$ are defined by
$$
     \hat{X}_\mu \equiv \lambda\ \hat{\Sigma}_{\mu a}
\eqn\eqD
$$
$$
     \hat{P}_\mu \equiv \hbar /R \ \hat{\Sigma}_{\mu b},   
\eqn\eqE
$$
together with $D$-dimensional angular momentum operator $\hat{M}_{\mu \nu}$
$$
   \hat{M}_{\mu \nu} \equiv \hbar \hat{\Sigma}_{\mu \nu}
\eqn\eqF
$$ 
and the so-called reciprocity operator
$$
    \hat{N}\equiv \lambda /R\ \hat{\Sigma}_{ab}.
\eqn\eqG
$$
In the above expressions, the so-called contraction-parameters $\lambda$ in \eqD\ and $R$ in 
\eqE\ are to be fundamental constants of YSTA, as will be seen below.  

Operators  $( \hat{X}_\mu, \hat{P}_\mu, \hat{M}_{\mu \nu}, \hat{N} )$ defined above 
constitute Yang's quantized space-time algebra, YSTA:
$$
   [ \hat{X}_\mu, \hat{X}_\nu ] = - i \lambda^2/\hbar \hat{M}_{\mu \nu}
\eqn\eqH
$$
$$
   [\hat{P}_\mu,\hat{P}_\nu ] = - i\hbar / R^2\ \hat{M}_{\mu \nu}
\eqn\eqI
$$
$$
      [\hat{X}_\mu, \hat{P}_\nu ] = - i \hbar \hat{N} \delta_{\mu \nu}
\eqn\eqJ
$$
$$
     [ \hat{N}, \hat{X}_\mu ] = - i \lambda^2 /\hbar  \hat{P}_\mu
\eqn\eqK
$$
$$
      [ \hat{N}, \hat{P}_\mu ] = -i \hbar/ R^2\ \hat{X}_\mu,
\eqn\eqL
$$
with familiar relations among ${\hat{M}_{\mu \nu}}\ 's$ omitted. 

At this point, it is important to notice the following simple fact that 
${\hat\Sigma}_{MN}$ with $M, N$ being the same metric signature have discrete eigenvalues and 
those with $M, N$ being opposite metric signature have continuous eigenvalues, as was shown 
explicitly in ref. \Ti. For the subsequent arguments, let us reconfirm this fact through 
space-time operators $\hat{X_\mu}$, 
$$
      \hat{X}_i = \lambda \hat{\Sigma}_{ia}= i\lambda (- q_i \partder{}{q_a}+q_a\partder{}{q_i}),
\eqn\eqM
$$
$$
     \hat{X}_0 = \lambda \hat{\Sigma}_{0a} = i \lambda (q_0 \partder{}{q_a}+ q_a \partder{}{q_0}).
\eqn\eqN
$$

They are rewritten as 
$$
\eqalign{
          &\hat{X}_i= \lambda {1 \over i} \partder{}{\alpha_i},  \cr
          & \hat{X}_0= \lambda {1 \over i} \partder{}{\alpha_0}
}
\eqn\eqO
$$
with $\alpha_i\ (0 \le |\alpha_i| \le \pi )$ and $\alpha_0\ (0 \le |\alpha_0| \le \infty)$ 
defined through   $q_i / q_a = \tan {\alpha}_i$ and  $q_0 / q_a = \tanh {\alpha}_0$, 
respectively. As was remarked above,  $\hat{X}_i$ and $\hat{X}_0$ in \eqO, respectively, have 
discrete and continuous eigenvalues, $ \lambda m_i$ with $\pm$ integer $m_i$ and real number 
$t$. The corresponding eigenfunctions are give by
$$
\eqalign{
    \ket{{\hat{\Sigma}}_{ia}= m_i} \sim \exp ({i\ m_i \alpha_i}) &= \exp [{i\ m_i\ {\tan}^{-1} 
( q_i / q_a)}] \cr
                          &= [( q_a - i q_i)/ ( q_a  +  iq_i)]^{i m_i/2}
}
\eqn\eqP
$$
and
$$
\eqalign{
    \ket{{\hat{\Sigma}}_{0a}= t/\lambda} \sim \exp [i\ (t/ \lambda )\ \alpha_0] &= \exp [i\ 
(t/\lambda)\ {\tanh}^{-1} (q_0 / q_a)] \cr
 &= [ ( q_a + q_0 ) / ( q_a - q_0 ) ]^{i t /(2 \lambda )},
}
\eqn\eqQ
$$
where normalization constants are omitted.

Needless to say, the eigenstate $\ket{{\hat{\Sigma}}_{0a}=t/\lambda}$ given by \eqQ\ is concerned
 with a boost operator ${\hat{\Sigma}}_{0a}$ of $S(D+1,1)$, and its continuous eigenvalue 
$t\ (/ \lambda)$ is to be identified with eigenvalue $t$ of time operator $\hat{X}_{0a}$, \eqN. 
This fact implies that Yang's space-time algebra (YSTA) presupposes for its representation space to take 
representation bases like $\ket{{\hat{\Sigma}}_{0a}= t/\lambda,n...}$, where $n...$  
denotes eigenvalues of maximal commuting set of compact subalgebra of $SO(D+1,1)$ which are 
commutative with ${\hat{\Sigma}}_{0a}$, for instance, ${\hat{\Sigma}}_{b1}$, 
${\hat{\Sigma}}_{23},$..., ${\hat{\Sigma}}_{89}$, when $D=11$.\footnote{*)}{It corresponds, 
in the case of unitary representation of Lorentz group $SO(3,1)$, to take $K_3\ (\sim \Sigma_{03})$ and $J_3\
 (\sim \Sigma_{12})$ to be diagonal, which have continuous and 
discrete eigenvalues, respectively, instead of ${\bf J}^2$ and $J_3$ in a regular 
representation.} 

Indeed, an infinite dimensional linear space expanded by $\ket{{\hat{\Sigma}}_{0a}= t/\lambda, 
n...}$ mentioned above provides a Hilbert space, called hereafter Hilbert space I according 
to ref.\Ti, which becomes a representation space of unitary infinite dimensional representation 
of $SO(D+1,1)$ and of YSTA. It is a kind of ``quasi-regular representation"\refmark\QR\ of SO(D+1,1), 
which is reducible to the familiar regular representations of $SO(D+1,1)$ algebra realized 
on its compact subalgebra. We expect the following expansion formula to hold  
$$
\ket{{\hat{\Sigma}}_{0a}= t/\lambda, n...}= \sum_{\sigma 's}\ \sum_{jm...}\  C^{\sigma, n... }_{jm...} 
(t/\lambda)\ \ket{\ \sigma 's ; j(j+1),m, ...},
\eqn\eqR
$$      
where $\ket{\sigma 's ; j(j+1), m,...}$ on the right hand side describe familiar unitary 
irreducible representation bases of $SO(D+1,1)$ with Casimir invariants $\sigma 's$, continuous 
or discrete in general, over which the summation $\Sigma_{\sigma's}$ ranges. Expansion 
coefficients $C^{\sigma 's}_{jm...} (t/\lambda)$ are to be calculated by making use of the functional 
form \eqQ\ for $\ket{{\hat{\Sigma}}_{0a}= t/\lambda, ...}$ on the left hand side.

Before closing this section, let us remark on the translation operation in YSTA, which was one of 
central motivation of YSTA beyond the original Snyder's quantized space-time algebra, where this 
operation did not work.  Let us define D-dimensional translation operator ${\hat T}$ with 
infinitesimal parameters $\alpha_\mu$ by
$$
          {\hat T}(\alpha_\mu) = \exp\ i\ (\alpha_\mu\ {\hat P}_\mu).
\eqn\eqS
$$

One finds that this operator induces infinitesimal transformation on $\hat{X}_\mu$ 
$$
             \hat{X}_\mu \rightarrow \hat{X}_\mu + \alpha_\mu\ \hat{N}
\eqn\eqT
$$
together with
$$
               \hat{N} \rightarrow \hat{N} -  \alpha_\mu\ \hat{X}_\mu / R^2.
\eqn\eqU
$$
This result is well understood, if one notices that the momentum operator ${\hat P}_\mu$ is 
nothing but generator of infinitesimal rotation on $b-\mu$ plane, ${\hat \Sigma}_{\mu b}$ given 
in \eqE.

However, let us here notice that the reciprocity operator $\hat{N} (= \lambda R^{-1} 
\hat{\Sigma}_{ab})$ defined in \eqG\ is an operator with discrete eigenvalues $ n\ 
(\lambda R^{-1})$, $n$ being $\pm$ integer and the displacement $ \alpha_\mu \hat{N}$ in \eqT\ is 
noncommutative with ${\hat X}_\mu$.  Therefore, it is important to see in what limit ordinary 
translations familiar in quantum mechanics may be exactly restored. Indeed, one finds them in a following 
limit of contraction-parameters, $\lambda$ and $R$,
$$
\eqalign{
      & \lambda \rightarrow 0  \cr
      & R \rightarrow \infty,
}
\eqn\eqV 
$$ 
in conformity with a condition
$$
     \hat N (= \lambda R^{-1} {\hat \Sigma}_{ab}) \rightarrow 1.
\eqn\eqW
$$

In fact, the condition \eqW\ under the limit \eqV\ necessitates a large limit of discrete 
eigenvalues of ${\hat\Sigma}_{ab}$\ in order for $\hat N$\ to survive with nonvanishing value $1$. 
Furthermore, one finds that the above limit completely restores the ordinary commutative 
space-time in addition to Heisenberg's uncertainty relation, as seen in \eqH\ to \eqL. 
This fact reminds us Bohr's correspondence principle at the birth of quantum mechanics, that is, 
quantum mechanics tends to classical mechanics in a large limit of quantum numbers.

\chapter{Noncommutative Field Theory on YSTA, Covariant Moyal Star Product and Matrix Model}

Now we are in a position to rewrite the matrix model\refmark\M\ in terms of noncommutative quantum 
field theory on YSTA. The so-called kinetic term and potential term of $D_0$ branes or D-particles 
are described in the following form
$$
               L = \alpha\ \Tr\ (P_i)^2 - \beta\ \Tr\ [ X_i, X_j ]^2 
 ,
\eqn\eqAA
$$
where $X_i$ and $P_i$ are, respectively, $ N \times N$ matrices of position and momentum of $N$ D-particles, 
and coefficient constants $\alpha$ and $\beta$ are given in terms of fundamental constants of string theory.

 At this 
point, we wish to rewrite the above Lagrangian covariantly by taking into consideration the 
following relation
$$
 {{\hat\Sigma}_{KL}}^2 = 2\ {{\hat\Sigma}_{\mu b}}^2 + {{\hat\Sigma}_{\mu\nu}}^2= 2 (R /\hbar)^2 
{{\hat P}_\mu}^2  - {\lambda}^{-4}\ [{\hat X}_\mu, {\hat X}_\nu]^2,
\eqn\eqAB
$$  
where ${{\hat\Sigma}_{KL}}^2$ denotes $(D+1)$-dimensional Casimir operator of $SO(D,1)$, with 
$K, L$ ranging over $\mu$ and $b$.

Indeed, in place of $L$ in \eqAA, we propose an action $\bar L$ of noncommutative field theory on 
YSTA by 
$$
\eqalign{
  {\bar L}
 &= A\ \Tr\ \bigl[\ [{\hat \Sigma}_{KL}, {\hat D}^\dagger]\ [{\hat \Sigma}_{KL}, {\hat D}]\ \bigr] 
\cr
& =A'\ \Tr\ \bigl[\ 2\ (R^2 /\hbar^2)\  [{\hat P}_\mu, {\hat D}^\dagger ]\ [ {\hat P}_\mu, \hat D] 
- {\lambda}^{-4 }\ [\ [{\hat X}_\mu, {\hat X}_\nu],\ {\hat D}^\dagger]\ [ [{\hat X}_\mu, 
{\hat X}_\nu], \hat D]\ \bigr],
}
\eqn\eqAC
$$
where $\hat D$ and ${\hat D}^\dagger$, respectively, denote $D_0$ brane field operator on YSTA and 
its hermitian conjugate.  These fields on YSTA must be also operators working on Hilbert space I discussed 
in section 2, that is, the representation space of ${\hat \Sigma}_{AB}$. In ref.\Ti, we studied the time 
development of D-field operator ${\hat D}$ under the assumption 
that in light-cone frame, ${\hat D}$  becomes diagonal with respect to light-cone time, or 
$ [X_+,  \hat D] = 0$.\footnote{*)}{In ref.\Ti, the potential term was regarded as interaction 
term in a naive sense, rewriting it as quartic term of $D_0$ brane field, while the argument for 
(light-cone) time development of field used there is still applicable in the present case.} 

In the present paper, however, in order to see a short distance behavior of the present 
noncommutative field theory, we investigate the field equation of $D_0$\ brane by making use of 
the familiar method of Moyal star product of noncommutative field theory stated in Introduction. 
At this point, it is important to note that in the present case, Moyal star product concerning 
Lorentz-covariant YSTA becomes also Lorentz-covariant. 

It is well known that if a system is described by $2n$ canonical variables, $q_A$ and $p_A$ 
$(A=1,2,...,n)$, or the corresponding quantized variables, ${\hat q}_A$ and ${\hat p_A}$, operator 
product of any two functions $ \hat F (\hat q,\hat p)$ and $\hat G(\hat q, \hat p)$ is accompanied 
with Moyal star product of the corresponding classical functions, $F(q,p)$ and $G(q,p)$, that is,
$$
\eqalign{
{\hat F} (\hat q,\hat p)\ {\hat G} (\hat q, \hat p)\ \sim\ & F (q, p) *  G(q, p) \cr 
&\equiv\ F ( q, p)\ \exp\  {i \over 2}\ (\  \overleftarrow{\partder{}{q_A}}\ \overrightarrow
{\partder{}{p_A}}  - \overleftarrow{\partder{}{p_A}}\ \overrightarrow{\partder{}{q_A}})\ 
G (q, p),
}
\eqn\eqAE
$$
when
$$
\eqalign{
{\hat F} (\hat q,\hat p) & \sim F(q, p), \cr
{\hat G} (\hat q, \hat p) &\sim G(q, p).
}
\eqn\eqAF     
$$  

In our present case, it is important to note that basic quantities of YSTA, 
${\hat \Sigma}_{MN}$ in \eqB, are expressed by
$$
{\hat \Sigma}_{MN} =  (- {\hat q}_M  {\hat p}_N + {\hat q}_N {\hat p}_M ),
\eqn\eqAG
$$
so the relation \eqAE\ turns to give the following covariant Moyal star product for any 
two functions, ${\hat F}$ and ${\hat G}$ on ${\hat \Sigma}(\hat q, \hat p)$,
$$
\eqalign{
{\hat F} (\hat \Sigma)\ {\hat G} (\hat \Sigma ) \sim & F (\Sigma) * G(\Sigma) \cr
& = F (\Sigma )\  \exp\  {i \over 2}\ (\  \overleftarrow{\partder{}{\Sigma_{MN}}}\ 
{\Sigma }_{NO}\ \overrightarrow{\partder{}{{\Sigma}_{OM}}})\ G (\Sigma),
}
\eqn\eqAF
$$
where ${\Sigma}_{MN} = ( - q_M p_N  + q_N p_M).$
 
Furthermore, by remarking that $[ \hat F, \hat G]$ in \eqAC\ is replaced by Moyal bracket, 
$\{F, G \}_M \equiv F * G -G * F $, we find
$$
\eqalign{
   [ {\hat \Sigma}_{KL} , {\hat D} ({\hat \Sigma}) ] &\sim \{ {\Sigma}_{KL}, D (\Sigma ) \}_M \cr
&= ( - {\Sigma}_{aK} {\partder{} {\Sigma}_{aL}} + {\Sigma}_{aL} {\partder{} {\Sigma}_{aK}} ) D 
(\Sigma).
}
\eqn\eqAG
$$
It should be noted that at the last step we assumed, self-consistently with $D_0$ brane field equation 
to be derived below, that $D_0$ brane field operator depends only on $\hat \Sigma_{aK}$, that is, 
$\hat X_\mu $ and $\hat N $, a minimum set which includes space-time variables and allows translation 
operations given by \eqT\ and \eqU.  

Applying variational principle on the classical action corresponding to
$ \bar L$, we obtain the field equation of $D_0$ brane
$$
[\ {\Sigma_{aK}}^2\ (\partder{} {\Sigma_{aL}})^2 - (\Sigma_{aK}  \partder{} {\Sigma_{aK}})^2 - 
(D-1) \Sigma_{aK}  \partder{} {\Sigma_{aK}}\ ]\ D (\Sigma_{aK}) = 0
\eqn\eqAH
$$
or
$$
\eqalign{
[( {X_\sigma}^2 + & R^2 N^2 ) \big( (\partder{} {X_\mu})^2 + R^{-2} (\partder{} {N})^2)\big)         \cr
 - & ( X_\mu \partder{} {X_\mu} + N \partder{}{N})^2 - (D-1) ( X_\mu \partder{} {X_\mu}+ 
N \partder{}{N}) ] D ( X, N) = 0,
}
{\eqAH}'
$$
with $D$ being the dimension of space-time $X_\mu$.

In this way, we have arrived at $D_0$ brane field equation with a nontrivial form beyond simple 
Klein-Gold equation, \eqAH\ or 
{\eqAH}', which reflects noncommutative space-time structure proper to YSTA. 
In fact, one finds that a simple Fourier analysis does not work well to find its solutions, on 
account of Lorentz-covariant couplings among individual degrees of freedom of space-time coming from their 
noncommutativity. Fourier transform of $D_0$ brane field  turns out  to satisfy entirely the same 
form of differential equation as \eqAH, reflecting a distinctive invariance of the latter field 
equation under exchange, ${\Sigma}_{aK} \leftrightarrow \partder{} {{\Sigma}_{aK}}$. 

Furthermore, one should notice that the field equation {\eqAH}'\ satisfies invariance under 
infinitesimal translations, $X_\mu \rightarrow X_\mu + \alpha_\mu N,\  N \rightarrow  N - 
\alpha_\mu X_\mu / R^2$ corresponding to \eqT\ and \eqU, and tends to a familiar massless 
Klein-Gordon equation in the contraction limit, \eqV\ under the condition \eqW, that is, in the 
quantum-mechanical limit of YSTA stated at the end in section 2. 

One can confirm by elementary calculations that the field equation admits, instead of familiar 
Fourier series with respect to individual coordinates $\Sigma_{aK}$, a series of Lorentz-invariant 
solutions with a continuous parameter $\alpha$ 
$$
\eqalign{
          &\exp\ i \alpha\ {\Sigma_{aK}}^2    \cr
          & = \exp\ i \alpha\  ( X_\mu^2 + R^2 N^2 ) /\lambda^2,
}
\eqn\eqAI
$$
whose existence turns out to reflect scale invariance of the field equation \eqAH\ under 
$ \Sigma_{aK} \rightarrow \rho\ \Sigma_{aK}$. Therefore, as a result of Fourier integral of the 
above series of solutions, one finds that any function of ${\Sigma_{aK}}^2$ can be a solution of 
the field equation. 

There exists a special solution
$$
\eqalign{
   (2\pi \lambda)^{-2} \int_{-\infty}^{+\infty} d\alpha\  \exp i \alpha\ 
( X_\mu^2 + R^2 N^2 ) /\lambda^2  &=  \delta\ (\ X_\mu^2 + R^2 N^2  )   \cr
& = \delta\ ( X_\mu^2 + \lambda^2 \Sigma_{ab}^2 ).
}
\eqn\eqAJ
$$
It reminds us of a regularized $D (x)$ function, ${\tilde D} (x)$, early proposed by Markov,
\Ref\Mar\markov
$$
\eqalign
{
{\tilde D} ( x ) &\equiv - \epsilon ( x_0 )\ /(2\pi)\ \delta ( {x_\mu}^2 + a^2 )  \cr
  & ( = \int d\kappa^2\ [\ \delta (\kappa^2) - {a\over{2\kappa}}\ J_1 (a \kappa)\ ]\ 
\Delta ( x; \kappa^2 )),
}
\eqn\eqAK
$$
where one finds that light-cone singularity of conventional $\Delta ( x; \kappa^2 )$ function is 
shifted to time-like region by a certain universal length $a$ corresponding to $\lambda$ in the 
present approach.

\def\markov{M.A.~Markov, Nucl.~Phy. {\bf 10} (1959) 140.}

\chapter{Concluding Remarks}

In this work we have studied a possible connection between noncommutative field theory on Yang's 
quantized space-time algebra (YSTA) and Matrix Model. We have found several important facts to 
support this possibility. First of all, we showed that the so-called kinetic term and potential 
term of D-particles of the matrix model are described covariantly and geometrically in terms 
of Casimir operator of $SO(D,1)$, a subalgebra of the primary algebra $SO(D+1,1)$ underlying YSTA, 
while they are usually understood to originate in Yang-Mills gauge theory.

In fact, it is well known that infinite-dimensional matrix representation of noncommutative 
position coordinates of D-particles in the matrix model has the origin in $U(N)$ Yang-Mills gauge 
symmetry caused by infinite $N$ D-particles.\ref\wi\ It should be emphasized, however, that M-theory as 
the matrix model of D-particles must be ultimately a master theory which rather underlies string theory 
and hence Yang-Mills gauge symmetry itself, because D-particles in the matrix model are regarded as 
fundamental constituents of strings.

\def\wi{E.~Witten, ``Bound States of Strings and $p$-Branes,"  Nucl.~Phys.~{\bf B460} (1995) 335, 
hep-th/9510135.}

In this connection, it is important to note that Hilbert space I defined in section 2, can be regarded 
as a representation space either of unitary infinite 
dimensional realization of YSTA or $U(N)$ gauge group with a large $N$ limit in the matrix model.
     
In section 3, with aim of seeking for a short distance behavior of the present noncommutative 
field theory on YSTA, we have derived the field equation of $D_0$ brane on YSTA with a nontrivial form, 
 \eqAH\ or {\eqAH}', 
by means of covariant Moyal star product \eqAF\ proper to YSTA,
which clearly reflects the quantized noncommutative space-time structure of YSTA. Indeed, to solve 
the equation, a simple Fourier analysis does not work well on 
account of couplings among individual degrees of freedom of noncommutative space-time. Fourier transform 
of $D_0$ brane field tends to satisfy entirely the same form of differential 
equation as \eqAH, reflecting a distinctive invariance of the latter field equation under exchanges, 
${\Sigma}_{aK} \leftrightarrow \partder{} {{\Sigma}_{aK}}$. The situation may be well understood 
also from the fact that Fourier transform of a special solution $\exp\ (i \alpha {\Sigma_{aK}}^2)$ 
given in \eqAI, that is, $F_\alpha (k) \equiv \int \cdots \int\ (d\Sigma)^{D+1} \exp\ (i k_M 
{\Sigma}_{aM}) \exp\ (i \alpha {\Sigma_{aK}}^2)$ is also given in a form  $\sim \exp -
(i {k_L}^2 /4\alpha).$  

Furthermore, we showed that the field equation {\eqAH}'\ has invariance under translations and 
scale transformation besides Lorentz-invariance and turns to a familiar massless Klein-Gordon 
equation in the quantum-mechanical limit of YSTA, \eqV\ and \eqW, in which commutative space-time, 
Heisenberg uncertainty relation, translations and so forth in the ordinary quantum mechanics are 
all restored. 

By preliminary considerations, we have found a series of Lorentz-covariant solutions with a 
continuous parameter $\alpha$, \eqAI, and a special invariant function \eqAJ, which 
resembles a modified $\Delta$ function, ${\tilde D}$, early proposed by Markov\refmark\Mar\ towards 
divergence-free quantum field theory. The latter function which is free from light-cone 
singularity is given in terms of superposition of a series of $\Delta(x, \kappa^2)$ functions of fields 
with continuous mass $\kappa$ and indefinite metric, as seen in \eqAK.
  
According to the matrix model, Lagrangian of $D_0$ branes \eqAA\ and hence our field equation 
\eqAH\ must already include interactions among $D_0$ branes except those through super-partners which were 
entirely neglected in the present paper.\footnote{*)}{Supersymmetric YSTA can be easily realized by upgrading $(D+2)$-dimensional parameter space $( q_M )$ 
satisfying \eqB\ to the so-called superspace $(q,\theta)$.} Extensive studies on our field equation \eqAH\ 
and its supersymmetric extension have a vital importance, but must be left in future.

\ack

The author would like to thank participants in Meeting on ``Mathematical Logic of Quantum System 
and its Application to Quantum Computer" for valuable discussions and suggestions, which was held 
at Research Institute for Mathematical Sciences, Kyoto University, in Feb. 2004.

\refout

\end